\documentclass[aps,prl,twocolumn,superscriptaddress,showpacs,preprintnumbers,amsmath,amssymb]{revtex4}

\usepackage{graphicx}
\usepackage{dcolumn}

\graphicspath{{ps}}

\begin{document}

\title{ \quad\\[1.0cm] Search for Invisible Decay of the $\Upsilon(1S)$}

\affiliation{Budker Institute of Nuclear Physics, Novosibirsk}
\affiliation{Chiba University, Chiba}
\affiliation{Chonnam National University, Kwangju}
\affiliation{University of Cincinnati, Cincinnati, Ohio 45221}
\affiliation{Department of Physics, Fu Jen Catholic University, Taipei}
\affiliation{Justus-Liebig-Universit\"at Gie\ss{}en, Gie\ss{}en}
\affiliation{The Graduate University for Advanced Studies, Hayama, Japan}
\affiliation{University of Hawaii, Honolulu, Hawaii 96822}
\affiliation{High Energy Accelerator Research Organization (KEK), Tsukuba}
\affiliation{University of Illinois at Urbana-Champaign, Urbana, Illinois 61801}
\affiliation{Institute of High Energy Physics, Chinese Academy of Sciences, Beijing}
\affiliation{Institute of High Energy Physics, Vienna}
\affiliation{Institute of High Energy Physics, Protvino}
\affiliation{Institute for Theoretical and Experimental Physics, Moscow}
\affiliation{J. Stefan Institute, Ljubljana}
\affiliation{Kanagawa University, Yokohama}
\affiliation{Korea University, Seoul}
\affiliation{Kyungpook National University, Taegu}
\affiliation{Swiss Federal Institute of Technology of Lausanne, EPFL, Lausanne}
\affiliation{University of Ljubljana, Ljubljana}
\affiliation{University of Maribor, Maribor}
\affiliation{University of Melbourne, Victoria}
\affiliation{Nagoya University, Nagoya}
\affiliation{Nara Women's University, Nara}
\affiliation{National Central University, Chung-li}
\affiliation{National United University, Miao Li}
\affiliation{Department of Physics, National Taiwan University, Taipei}
\affiliation{H. Niewodniczanski Institute of Nuclear Physics, Krakow}
\affiliation{Nippon Dental University, Niigata}
\affiliation{Niigata University, Niigata}
\affiliation{University of Nova Gorica, Nova Gorica}
\affiliation{Osaka City University, Osaka}
\affiliation{Osaka University, Osaka}
\affiliation{Panjab University, Chandigarh}
\affiliation{Peking University, Beijing}
\affiliation{University of Pittsburgh, Pittsburgh, Pennsylvania 15260}
\affiliation{RIKEN BNL Research Center, Upton, New York 11973}
\affiliation{Saga University, Saga}
\affiliation{University of Science and Technology of China, Hefei}
\affiliation{Seoul National University, Seoul}
\affiliation{Shinshu University, Nagano}
\affiliation{Sungkyunkwan University, Suwon}
\affiliation{University of Sydney, Sydney NSW}
\affiliation{Tata Institute of Fundamental Research, Bombay}
\affiliation{Toho University, Funabashi}
\affiliation{Tohoku Gakuin University, Tagajo}
\affiliation{Tohoku University, Sendai}
\affiliation{Department of Physics, University of Tokyo, Tokyo}
\affiliation{Tokyo Institute of Technology, Tokyo}
\affiliation{Tokyo Metropolitan University, Tokyo}
\affiliation{Tokyo University of Agriculture and Technology, Tokyo}
\affiliation{Virginia Polytechnic Institute and State University, Blacksburg, Virginia 24061}
\affiliation{Yonsei University, Seoul}
  \author{O.~Tajima}\affiliation{High Energy Accelerator Research Organization (KEK), Tsukuba} 
  \author{H.~Hayashii}\affiliation{Nara Women's University, Nara} 
  \author{M.~Hazumi}\affiliation{High Energy Accelerator Research Organization (KEK), Tsukuba} 
  \author{K.~Inami}\affiliation{Nagoya University, Nagoya} 
  \author{Y.~Iwasaki}\affiliation{High Energy Accelerator Research Organization (KEK), Tsukuba} 
  \author{S.~Uehara}\affiliation{High Energy Accelerator Research Organization (KEK), Tsukuba} 
  \author{K.~Abe}\affiliation{High Energy Accelerator Research Organization (KEK), Tsukuba} 
  \author{I.~Adachi}\affiliation{High Energy Accelerator Research Organization (KEK), Tsukuba} 
  \author{H.~Aihara}\affiliation{Department of Physics, University of Tokyo, Tokyo} 
  \author{D.~Anipko}\affiliation{Budker Institute of Nuclear Physics, Novosibirsk} 
% ...
\collaboration{The Belle Collaboration}

\begin{abstract}

We report results of a search for the invisible decay of the $\Upsilon(1S)$ 
via the $\Upsilon(3S) \to \pi^+\pi^-\Upsilon(1S)$ transition
using a data sample of 
$2.9\,{\rm fb}^{-1}$ at the $\Upsilon(3S)$ resonance.
The data were collected with the Belle detector 
at the KEKB asymmetric-energy $e^+ e^-$ collider.
No signal is found, and an upper limit for the branching fraction
at the 90\% confidence level is determined to be
${\cal B} (\Upsilon(1S) \to {\rm invisible})$ $<$ 2.5 $\times$ $10^{-3}$.

\end{abstract}


\maketitle

\tighten

{\renewcommand{\thefootnote}{\fnsymbol{footnote}}}
\setcounter{footnote}{0}

An invisible particle decay mode is defined as one where the
final state particles interact so weakly that they are not observable in
a detector.
The standard model (SM) predicts that there are 
no invisible particles except neutrinos.
If the invisible decay rate is observed to have a larger branching fraction 
than the SM prediction, 
it implies physics beyond the SM.
One possibility is a decay into dark matter particles, $\chi$, 
whose existence is strongly suggested by several astronomical
observations~\cite{bib_dm_evidence}.
In the SM, quarkonium decay to the two-neutrino final
state is predicted to have a branching fraction
${\cal B}(\Upsilon(1S) \to \nu\bar{\nu})$ 
= $(9.9 \pm 0.5) \times 10^{-6}$
\cite{bib_upsilon2nunubar}.
A much larger invisible branching fraction
${\cal B}(\Upsilon(1S)  \to \chi \chi) \simeq 6 \times 10^{-3}$
is predicted \cite{bib_BobMcElrath_invisble_upsilon}
for dark matter particles that are lighter than the $b$ quark.
Here, the pair annihilation cross section of dark matter particles
to a SM quark pair 
$\sigma(\chi \chi \to q \bar q)$
is estimated based on cosmological arguments, 
and the time-reversed reaction is assumed to have the same cross section,
i.e. $\sigma(q \bar q\to \chi \chi) = \sigma(\chi \chi \to q \bar q)$.
The previous upper limits for the invisible $\Upsilon(1S)$ branching fraction
were reported by 
ARGUS (23 $\times$ $10^{-3}$ with 90\% confidence level)
\cite{bib_argus}
and 
CLEO (50 $\times$ $10^{-3}$ with 95\% confidence level)
\cite{bib_cleo}, 
which are about one order of magnitude above 
the prediction \cite{bib_BobMcElrath_invisble_upsilon}.

In this letter, we present the result of a search for 
the invisible decay of the $\Upsilon(1S)$.
The data sample used consists of 
$2.9\,{\rm fb}^{-1}$ collected on the $\Upsilon(3S)$ resonance
(11 $\times$ $10^6$ $\Upsilon(3S)$)
 with the Belle detector~\cite{Belle}
at the KEKB asymmetric-energy
$e^+e^-$ (3.4 on 7.8~GeV on $\Upsilon(3S)$) collider~\cite{KEKB}.
The Belle detector is a large-solid-angle magnetic
spectrometer that
consists of 
a silicon vertex detector,
a central drift chamber (CDC), an array of
aerogel threshold \v{C}erenkov counters (ACC), 
a barrel-like arrangement of time-of-flight
scintillation counters (TOF), and an electromagnetic calorimeter
comprised of CsI(Tl) crystals (ECL) located inside 
a superconducting solenoid coil that provides a 1.5~T
magnetic field.  An iron flux-return located outside 
the coil is instrumented to detect $K_L^0$ mesons and to identify
muons (KLM).

For this search, we use the $\Upsilon(3S) \to \pi^+\pi^-\Upsilon(1S)$
decay where only the cascade $\pi^+ \pi^-$ pair is detected.
If an $\Upsilon(1S)$ decay into an invisible final state does occur, 
it would appear as a peak at the $\Upsilon(1S)$ mass (9.46
GeV/$c^2$) in the distribution of recoil mass against the 
$\pi^+ \pi^-$ system 
($M_{\pi^+\pi^-}^{\rm recoil}$)
without any detected decay products from the $\Upsilon(1S)$.
To provide a clean sample of $\Upsilon(1S)$ decays,
we choose to run at the $\Upsilon(3S)$ resonance.
Although the
$\Upsilon(2S) \to \pi^+\pi^-\Upsilon(1S)$
branching fraction is about four times larger, 
the low energy of the cascade pions results in a low trigger efficiency.
The effective cross section on the 
$\Upsilon(4S)$ resonance, 
$\sigma(e^+e^- \to \Upsilon(4S) \to \pi^+\pi^-\Upsilon(1S))$,
is much smaller than that at the $\Upsilon(3S)$
resonance ($\sim$ 1/1,000 \cite{bib_4spipi}).
To understand the reconstruction efficiency,
it is essential to check that our Monte Carlo (MC) simulation reproduces well
the properties of the $\Upsilon(3S) \to \pi^+\pi^-\Upsilon(1S)$ transition.
This is confirmed by using a control sample,
$\Upsilon(3S) \to \pi^+\pi^-\Upsilon(1S)$
followed by
$\Upsilon(1S) \to \mu^+\mu^-$.

In our trigger system~\cite{bib_trigger}, a charged track is 
required to have hits
in more than half the layers of the CDC.
Tracks that reach the outermost layer of the CDC are 
called ``long tracks''. 
To suppress beam-induced background events,
the trigger system requires
two or more charged tracks in an event,
and at least one of them should be a long track.
The trigger system also requires that the opening angle of the
two tracks in the transverse plane be larger than a certain
value for the purpose of identifying tracks being different
particles.
Careful evaluation of the trigger efficiency is important for this analysis.
The signal events contain only two charged tracks with relatively small
transverse momentum ($p_t$);
the available energy for the signal is 0.89 GeV,
which is shared between the two pions.
To evaluate the trigger efficiency, 
we took 
1.7 ${\rm fb}^{-1}$
of data with a trigger condition that required
a single long track in an event with a 1/500 pre-scale rate.

Charged tracks that are reconstructed in the CDC are required to
originate from the interaction point: 
the nearest approach of the trajectory to the collision point is
required to satisfy $dr$ $<$ 1 cm and $|dz|$ $<$ 3 cm,
where $dz$  is measured along the direction opposite to the positron beam
and $dr$ in the plane perpendicular to it.
We also require that the polar angle of the reconstructed track be within the detector acceptance,
 $17^{\circ}$ $<$ $\theta$ $<$ $150^{\circ}$.
Charged kaons are distinguished from pions 
based on TOF, ACC and CDC $dE/dx$ measurements.
Electron candidates are identified based on the ratio of the energy
detected in the ECL to the track momentum, the ECL shower shape, the
energy loss in the CDC and the response of the ACC.
Identification of muons is based on track
penetration depth and the hit pattern in the KLM system.

\begin{figure}[t]
\includegraphics[width=0.45\textwidth]{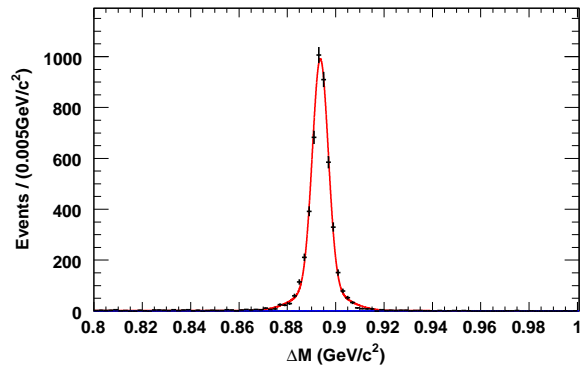}
\caption{
Mass difference 
$\Delta M$ = $M_{\mu^+\mu^-\pi^+\pi^-}-M_{\mu^+\mu^-}$
distribution where 
$M_{\mu^+\mu^-}$ lies in the $\Upsilon(1S)$ mass region.
The solid curve shows the fits to signal plus background distribution.
}
 \label{fig_upsilon_1s_mumu_dmassfit_dat}
\end{figure}

\begin{figure}[t]
\includegraphics[width=0.45\textwidth]{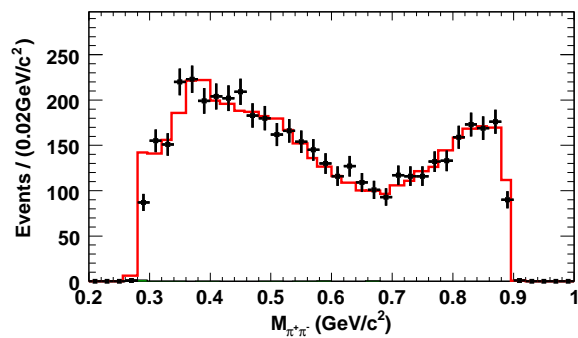}
 \caption{
 Invariant mass distribution of the two cascade pions for the 
 $\Upsilon(3S) \to \pi^+\pi^-\Upsilon(1S)$,
 $\Upsilon(1S) \to \mu^+\mu^-$ cascade decay candidates.
 The open histogram shows the same distribution for MC events.
 }
 \label{fig_upsilon_1s_mumu_mpipi}
\end{figure}

To reconstruct events for the control sample, 
we require four charged tracks in an event, 
$\mu^+$, $\mu^-$, $\pi^+$, and $\pi^-$.
The $\Upsilon(1S) \to \mu^+\mu^-$ candidates are selected 
by requiring 9.2 GeV/$c^2$ $<$ $M_{\mu^+\mu^-}$ $<$ 9.7 GeV/$c^2$
with muon identification for both tracks.
For the pions, we reject tracks that are positively identified as
electrons or muons.
Figure \ref{fig_upsilon_1s_mumu_dmassfit_dat} shows the distribution of
the mass difference 
$\Delta M$ = $M_{\mu^+\mu^-\pi^+\pi^-}-M_{\mu^+\mu^-}$.
The signal yield in the control sample is extracted from
an unbinned maximum likelihood fit in $\Delta M$. 
The signal shape is modeled with a triple Gaussian
while 
the background shape is modeled with a first order polynomial
whose slope is floated in the fit.
We obtain 4902 $\pm$ 71 (87 $\pm$ 15) signal (background) events 
in the range of 0.65 GeV/$c^2$ $<$ $\Delta M$ $<$ 1.15 GeV/$c^2$.
Using a detection efficiency of 39.7\% for the control sample, 
which is determined from MC calculations,
we estimate that 
498 $\times$ $10^3$ 
$\Upsilon(3S) \to \pi^+\pi^-\Upsilon(1S)$ 
decays are present in our data set.
We compare properties of the 
$\Upsilon(3S) \to \pi^+\pi^- \Upsilon(1S)$
decay with MC
using events in the region of 
0.875 GeV/$c^2$ $<$ $\Delta M$ $<$ 0.910 GeV/$c^2$,
where the signal purity is 99.9\%.
The distribution of $\pi^+\pi^-$ invariant masses,
$M_{\pi^+\pi^-}$ is
well reproduced by the MC as shown in Fig. \ref{fig_upsilon_1s_mumu_mpipi},
and the shape is
consistent with CLEO results \cite{bib_CLEO_3S_property}
and theoretical models \cite{bib_Mooxhay}.
We also confirm that the MC reproduces well the distributions for other
variables in the data such as 
$p_t$ of pion tracks,
the opening angle of the pions in the transverse plane ($\varphi_{\pi\pi}$),
the transverse momentum of the $\pi^+\pi^-$ system ($p_t^{\pi\pi}$),
the polar angle of the $\pi^+\pi^-$ system ($\cos \theta_{\pi\pi}$), 
and the polar angle of the muons.


For the selection of the invisible decay candidates, 
we require two oppositely charged tracks in the event, 
i.e.\ $\pi^+\pi^-$ tracks.
The total visible energy in the ECL is required to be less than 3 GeV
to reject $\Upsilon(1S)$ decays into
 final states consisting of neutral particles.
To minimize any possible trigger bias, 
we require $\varphi_{\pi\pi}$ $>$ $30^{\circ}$, $p_t$ $>$ 0.17 GeV/$c$ for the tracks, 
and $p_t$ $>$ 0.30 GeV/$c$ for at least one of the tracks.
Most combinatorial background is from the two-photon
processes 
$e^+ e^- \to e^+ e^- X$,
where the incident $e^+$ and $e^-$ escape detection 
and $X \to \pi^+\pi^-, \mu^+\mu^-, \pi^+\pi^-\pi^0$ or other states.
We reject tracks that are positively identified as
electrons, muons or kaons.
We require that 
no $\pi^0$ candidates are observed
in the event; 
we form $\pi^0$ candidates using pairs of photons 
having energies greater than 20 MeV,
and requiring a pair invariant mass within
$\pm16$ MeV/$c^2$ ($\sim3\sigma$) of the $\pi^0$ mass.
Events from two-photon process are typically boosted along the beam
direction so that the vector $p_t$ sum tends to be small. 
For further background suppression,
a Fisher discriminant variable ${\cal F}$ is constructed with the
linear combination of $|\cos\theta_{\pi\pi}|$, $p_t^{\pi\pi}$ 
and the maximum energy of the $\gamma$ candidates in the event 
($E_{\gamma}^{\rm max}$),
${\cal F} = 0.87 \times |\cos\theta_{\pi\pi}|
            -2.4 \times p_t^{\pi\pi}
	    + 1.43 \times E_{\gamma}^{\rm max}
$,
where coefficients
are optimized by using MC signal events
and background events in the $\pi^+\pi^-$ recoil mass sideband
(9.42 GeV/$c^2$ $<$ $M_{\pi^+\pi^-}^{\rm recoil}$ $<$ 9.44 GeV/$c^2$ or 
9.48 GeV/$c^2$ $<$ $M_{\pi^+\pi^-}^{\rm recoil}$ $<$ 9.50 GeV/$c^2$).
We require ${\cal F}$ $<$ $-$0.7;
this requirement is determined using the figure-of-merit, $S/\sqrt{B}$, 
where $S$ is the number of signal events in MC
and $B$ is the number of background events 
as mentioned above.
The overall detection efficiency, 
which is the product of 
the event reconstruction efficiency (9.1\%)
and 
the trigger efficiency (89.8\%),
is 8.2\%; 
thus we expect
244 signal events in our data sample for 
${\cal B}(\Upsilon(1S) \to \chi \chi)$ = 6 $\times$
$10^{-3}$.
The event reconstruction efficiency 
is obtained from MC. 

We estimate the trigger efficiency for signal events by
selecting two pion candidates
in the single track trigger data
under the same condition as for the signal.
We examine the trigger bits that are activated in each of the selected events.
The track-finding efficiency of the trigger system 
is the ratio of number of signal candidates 
that satisfy
the two-track trigger requirement to 
those that satisfy
the single track trigger requirement.
We evaluate the track-finding efficiency as a function of $p_t$ 
requiring a large opening angle 
$\varphi_{\pi\pi}$ $>$ $70^{\circ}$
(the possible bias for two-track separation is negligible in this case).
The efficiency for finding the long track is evaluated 
with a further requirement that both tracks are long tracks.
The efficiency of the long track finding is 97.1\% for $p_t$ $>$ 0.30 GeV/$c$.
We also evaluate the two-track separation efficiency as a function
of $\varphi_{\pi\pi}$, where we require $p_t$ $>$ 0.30 GeV/$c$ for
the long track and $p_t$ $>$ 0.17 GeV/$c$ for the other track.
The efficiency for the $\varphi_{\pi\pi}$ $>$ $30^{\circ}$ requirement is 92.5\%.
The detection efficiency for the two-pion signal in the trigger system, 
which is the product of 97.1\% and 92.5\%, is 89.8\%.

The systematic uncertainties are summarized in
Table \ref{table_syst}.
The uncertainty related to the track selection
is estimated from the difference of the 
signal yield in 
data and MC for the control sample 
when the $dr$, $dz$, particle identification
and other requirements are varied.
The error for the $\pi^0$ veto, Fisher discriminant, and other
requirements are also estimated in a similar way.
Uncertainties associated with the modeling of the 
$\Upsilon(3S) \to \pi^+\pi^-\Upsilon(1S)$ decay process are estimated
by distorting the $M_{\pi^+\pi^-}$ distribution within the statistical error; 
possible differences in other distributions are also considered.
The error for the trigger efficiency is obtained from the quadratic sum
of uncertainties for the track-finding efficiency and 
the two-track separation efficiency.
They are conservatively assigned as
the differences from 100\% efficiencies.
Various deviations are evaluated by changing the minimum $p_t$ criteria or
polar angle dependences. 
They are negligibly small.

\begin{table}[t]
\caption{Systematic uncertainties for
${\cal B}( \Upsilon(1S) \to {\rm invisible} )$
except for that from peaking background.
}
\label{table_syst}
\begin{tabular}
{@{\hspace{0.5cm}}l@{\hspace{0.5cm}} @{\hspace{0.5cm}}r@{\hspace{0.5cm}}}
\hline \hline
Source & (\%) \\
\hline
Track selection  & 5.6 \\
$\pi^0$ veto & 2.4 \\
Fisher discriminant & 6.1 \\
Other selection requirements & 1.1 \\
$\Upsilon(3S) \to \pi^+\pi^-\Upsilon(1S)$
& 7.6 \\
Trigger efficiency & 8.7 \\
Fit bias & 0.2 \\
Statistics of control sample & 1.4 \\
${\cal B}(\Upsilon \to \mu^+\mu^-)$ & 2.0 \\
\hline
Total & 14.7 \\
\hline \hline
\end{tabular}
\end{table}

\begin{table}[b]
\caption{
 Expected number of peaking background events.
}
\label{table_pbkg}
\begin{tabular}
{@{\hspace{0.5cm}}l@{\hspace{0.5cm}} @{\hspace{0.5cm}}r  c r@{\hspace{0.5cm}}}
\hline \hline
$\Upsilon(1S) \to \nu\bar{\nu}$ &  0.4 & $\pm$ & 0.1 \\
$\Upsilon(1S) \to \mu^+\mu^-$   & 77.3 & $\pm$ & 12.0 \\
$\Upsilon(1S) \to e^+e^-$     & 50.3 & $\pm$ &  8.2 \\
$\Upsilon(1S) \to \tau^+\tau^-$ &  5.2 & $\pm$ &  1.0 \\
Other $\Upsilon(1S)$  decay modes &  0.0 & $+$ &  2.8 \\
Other possible contributions &  0.0 & $+$ &  12.9 \\
\hline
Total & 133.2 & & $^{+19.7}_{-14.6}$\\
\hline \hline
\end{tabular}
\end{table}

For those cases where all of the $\Upsilon(1S)$  decay products go outside
of the detector acceptance, the $M_{\pi^+\pi^-}^{\rm recoil}$
distribution still peaks at the $\Upsilon(1S)$ mass and becomes
a background to the invisible decay signal. 
The largest sources of this ``peaking background'' 
are from decays to oppositely charged pairs such as 
$\Upsilon(1S) \to \mu^+\mu^-$
or 
$e^+e^-$,
where the two tracks tend to be back-to-back, so that  
when one track escapes into the forward acceptance hole, 
the other track tends to escape into the backward acceptance hole. 
The estimated contributions from different decay modes, based on MC,
are summarized in 
Table \ref{table_pbkg}.
The total number of peaking background events is 
133.2 $^{+19.7}_{-14.6}$,
where both statistical and systematic errors are included.

The systematic errors for the peaking background estimation can come
from uncertainties in the detection efficiency, branching fractions,
MC statistics and the track-finding efficiency for the tracks from the
$\Upsilon(1S)$ decay.
For the contribution from the track-finding efficiency, 
we evaluate the polar angle dependence of the track-finding efficiency
by using the control sample data.
Even if we select only one muon and two pions, 
we can calculate the momentum of the other muon track.
We compare the data and corresponding MC.
The amount of the peaking background differs by 3.5\% between data and MC.
The peaking background due to 
$\Upsilon(1S) \to e^+e^-$
is somewhat lower
than that due to 
$\Upsilon(1S) \to \mu^+\mu^-$.
This is due to the fact that the polar angle acceptance of the
ECL ($12.4^{\circ}$ -- $155.1^{\circ}$)
is larger than that of the CDC ($17.0^{\circ}$ -- $150.0^{\circ}$)
and 
large ECL energy deposits from an electrons can
be present in the event, 
and can veto the event.
Other two-body decays, such as 
$\Upsilon(1S) \to \pi^+\pi^-$
and 
$p\bar{p}$, 
are not observed,
and their contributions are included systematic error
using upper limits from the PDG \cite{bib_PDG}. 
We do not observe any background contribution 
in MC for other 
$\Upsilon(1S)$, 
$\Upsilon(3S)$ decays,
or modes
originating from initial state radiation.
The upper limits corresponding to MC statistics
are assigned as uncertainties.

\begin{figure}[t]
\includegraphics[width=0.45\textwidth]{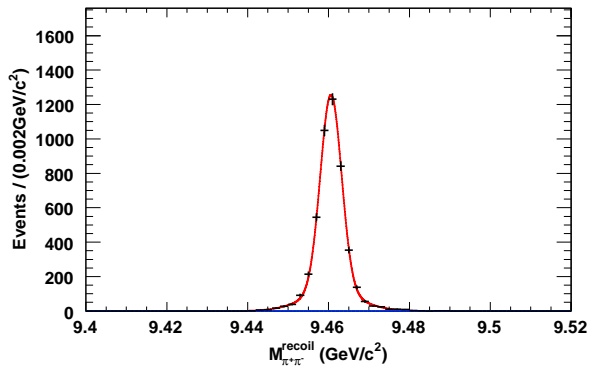}
 \caption{
 $M_{\pi^+\pi^-}^{\rm recoil}$ distribution for 
 the control sample
 $\Upsilon(3S) \to \pi^+\pi^-\Upsilon(1S)$, 
 $\Upsilon(1S) \to \mu^+\mu^-$ decay candidates.
 The solid curve shows the fit results.
 }
 \label{fig_invisiblefit_upsmumu}
\end{figure}

\begin{figure}[t]
\includegraphics[width=0.45\textwidth]{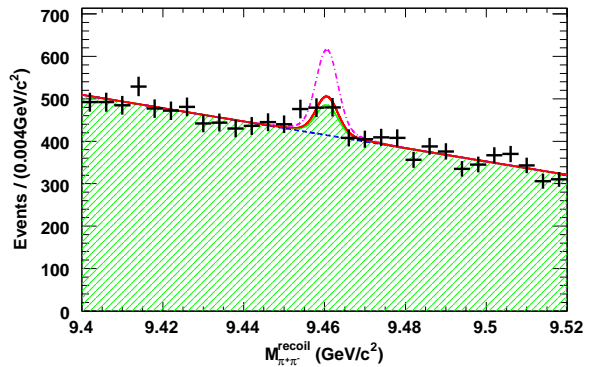}
 \caption{
 Recoil mass distribution against two pions, $M_{\pi^+\pi^-}^{\rm recoil}$.
 The solid curve shows the result of the fit to signal plus background distributions,
 shaded area shows the total background contribution, 
 dashed line shows the combinatorial background contribution,
 and the dot-dashed line shows the expected signal for
 ${\cal B}(\Upsilon(1S)  \to \chi \chi) = 6 \times 10^{-3}$.
 }
 \label{fig_invisiblefit_result}
\end{figure}

The signal extraction is performed by an unbinned extended maximum
likelihood fit to the $M_{\pi^+\pi^-}^{\rm recoil}$ 
distribution in the range
 9.40 GeV/$c^2$ $<$ $M_{\pi^+\pi^-}^{\rm recoil}$ $<$ 9.52 GeV/$c^2$.
The signal shape is modeled with a double Gaussian 
that is calibrated using the control sample
shown in Fig. \ref{fig_invisiblefit_upsmumu}.
In the fit, the amount of peaking background is fixed at the estimated
value and the same shape as the signal is used. 
The shape of the combinatorial background
is modeled with a first order polynomial
whose slope is floated.
Figure \ref{fig_invisiblefit_result} shows the 
$M_{\pi^+\pi^-}^{\rm recoil}$ distribution.
The extracted signal yield, 
$38 \pm 39$ events,
is consistent with zero observed events.
A $\chi^2$ test is performed for the $M_{\pi^+\pi^-}^{\rm recoil}$
distribution; we obtain $\chi^2/ndf$ = 23.4/27.
The upper limit for the branching fraction is determined
by the Feldman-Cousins frequentist approach \cite{bib_FeldmanCousins},
taking into account both statistical and systematic uncertainties.
We obtain 
${\cal B}( \Upsilon(1S) \to {\rm invisible} )$
$<$ 2.5 $\times$ $10^{-3}$ at the 90\% confidence level.

In summary, 
a search for the invisible decay of the $\Upsilon(1S)$ was performed
via the $\Upsilon(3S) \to \pi^+\pi^-\Upsilon(1S)$ transition.
In a $2.9\,{\rm fb}^{-1}$ data sample taken at the $\Upsilon(3S)$ resonance,
no signals were found. 
We obtain an upper limit of 
2.5 $\times$ $10^{-3}$ at the 90\% confidence level
for ${\cal B} (\Upsilon(1S) \to {\rm invisible})$.
This result disfavors the prediction 
in Ref.~\cite{bib_BobMcElrath_invisble_upsilon}
for the $\Upsilon(1S)$ decay to a pair
of dark matter particles that are lighter than the $b$ quark.

We thank the KEKB group for excellent operation of the
accelerator, the KEK cryogenics group for efficient solenoid
operations, and the KEK computer group and
the NII for valuable computing and Super-SINET network
support.  We acknowledge support from MEXT and JSPS (Japan);
ARC and DEST (Australia); NSFC and KIP of CAS (China); 
DST (India); MOEHRD, KOSEF and KRF (Korea); 
KBN (Poland); MIST (Russia); ARRS (Slovenia); SNSF (Switzerland); 
NSC and MOE (Taiwan); and DOE (USA).

\end{document}